\documentclass[sigconf,anonymous=false,review=false]{acmart}
\AtBeginDocument{%
  }

\setcopyright{acmlicensed}
\copyrightyear{2026}
\acmYear{2026}
\acmDOI{XXXXXXX.XXXXXXX}
\acmConference[SIGIR'49]{International ACM SIGIR Conference on Research and Development in Information Retrieval}{20–24 July 2026}{Melbourne | Naarm, Australia}
\acmISBN{978-1-4503-XXXX-X/2018/06}

\author{Moncef Garouani}
\email{Moncef.Garouani@irit.fr}
\orcid{0000-0003-2528-441X}
\affiliation{%
  \institution{IRIT, UMR5505 CNRS\\ Université Toulouse Capitole}
  \city{Toulouse}
  \country{France}
}

\author{Josiane Mothe}
\email{Josiane.Mothe@irit.fr}
\orcid{0000-0001-9273-2193}
\affiliation{%
\institution{IRIT, UMR5505 CNRS\\ UT2J, Université de Toulouse}
  \city{Toulouse}
  \country{France}
}

\renewcommand{\shortauthors}{Garouani et al.}

\usepackage{graphicx}
\usepackage{multirow}
\usepackage{booktabs}
\usepackage{tcolorbox}
\usepackage{listings}
\usepackage{xcolor}
\usepackage{graphicx}%
\usepackage{url}
\usepackage{hyperref}
\usepackage[table]{xcolor}
\usepackage{amsmath}

\definecolor{codegreen}{rgb}{0,0.6,0}
\definecolor{codegray}{rgb}{0.5,0.5,0.5}
\definecolor{codepurple}{rgb}{0.58,0,0.82}
\definecolor{backcolour}{rgb}{0.95,0.95,0.92}

\lstdefinestyle{mystyle}{
  backgroundcolor=\color{backcolour}, commentstyle=\color{codegreen},
  keywordstyle=\color{magenta},
  numberstyle=\tiny\color{codegray},
  stringstyle=\color{codepurple},
  basicstyle=\ttfamily\footnotesize,
  breakatwhitespace=true,         
  breaklines=true,                 
  captionpos=b,                    
  keepspaces=true,                 
  numbersep=5pt,                  
  showspaces=false,                
  showstringspaces=false,
  showtabs=false,                  
  tabsize=10
}

\begin{document}

\title[Improving Neural Retrieval with Attribution-Guided Query Rewriting]{Improving Neural Retrieval with Attribution-Guided \\Query Rewriting}




\begin{abstract}
Neural retrievers are effective but brittle: underspecified or ambiguous queries can misdirect ranking even when relevant documents exist. Existing approaches address this brittleness only partially: LLMs rewrite queries without retriever feedback, and explainability methods identify misleading tokens but are used for post-hoc analysis.
We close this loop and propose an attribution-guided query rewriting method that uses token-level explanations to guide query rewriting. For each query, we compute gradient-based token attributions from the retriever and then use these scores as soft guidance in a structured prompt to an LLM that clarifies weak or misleading query components while preserving intent. Evaluated on BEIR collections, the resulting rewrites consistently improve retrieval effectiveness over strong baselines, with larger gains for implicit or ambiguous information needs.
\end{abstract}




\maketitle

\section{Introduction}

Neural retrieval models (dense retrievers, cross-encoders, and hybrid architectures) have become central to information retrieval (IR), often  outperforming purely lexical approaches by capturing semantic similarity beyond exact term overlap \cite{6063-7_38,Thakur2024,zhao2024dense}.  Despite these advances, their effectiveness can drop sharply when user queries are underspecified, ambiguous, contain typos, or express an information need only implicitly ~\cite{hagen2024revisiting,sidiropoulos2024improving,li2025agent4ranking}. 
In such cases, the retriever may focus on unhelpful terms, ignore key constraints, or fail to reflect the user intent in the top-ranked results, even when relevant documents exist.

Query reformulation is a long-standing way to address these failures, from classical feedback and expansion techniques ~\cite{robertson1990term,harman1992relevance,rieh2006analysis,carpineto2012survey,esposito2020hybrid} to recent neural and LLM-based query rewriting~\cite{jagerman2023queryexpansionpromptinglarge,wang2023query2doc,xia2025knowledge}.

LLMs in particular can generate fluent paraphrases and expansions for short queries. However, most rewriting methods remain retriever-agnostic: they optimize for linguistic plausibility or semantic similarity without using signals from the downstream retriever. As a result, their impact on retrieval effectiveness is often inconsistent and may even be detrimental by introducing spurious or noisy terms~\cite{khattab2020colbert}. In parallel, explainable IR has developed token-level analysis tools (e.g., gradient- and perturbation-based attributions) to identify which query parts influence ranking decisions, but these techniques are typically used for post-hoc inspection rather than to improve retrieval \cite{11272475, han2021explaining,ribeiro2016should,sundararajan2017axiomatic}.

We bridge these two research directions by using token-level attributions as guidance for query rewriting. Given an input query, we first compute the contribution score for each query token from the retriever (via gradient-based attribution over the top retrieved documents). We then provide the full list of tokens and their scores to an LLM as soft guidance, encouraging it to preserve strongly helpful tokens and to clarify weak or misleading ones while preserving the original query intent. The rewritten query is finally re-issued to the same retriever in a one-shot closed loop. This design keeps the retriever unchanged and requires no additional training data or relevance judgments, making it easy to apply across retrieval architectures.


\sloppy

We evaluate our approach on standard IR collections from BEIR~\cite{thakur2021beir}, using representative retrievers from two neural retrieval families: SPLADE~\cite{spladev2} and TCT-ColBERT\cite{tctcolbert}. Guided query rewriting is performed using the Mistral LLM~\footnote{\url{https://huggingface.co/mistralai/Mistral-7B-Instruct-v0.3}}.  
Our experimental results show that attribution-guided query rewriting consistently improves retrieval effectiveness compared to both original queries and LLM-based rewriting methods that do not exploit attribution signals. We observe particularly strong gains for queries with implicit information needs and those involving domain-specific or ambiguous terminology. Ablation studies further demonstrate that attribution-based token selection leads to more effective rewrites than state-of-the-art strategies, highlighting the importance of explanation-driven control in the rewriting process.

\newpage
In summary, we contribute\,:
\begin{itemize}
    \item A retriever-aware query rewriting method that uses token-level attributions to guide edits. 
    \item A two-stage approach combining retriever-derived attribution signals with LLM-based, context-aware query reformulation.
    \item Empirical evidence on BEIR collections that attribution-guided rewriting outperforms strong baselines, including LLM-only rewriting and token-pruning variants.
    \item Evidences that explanations can play a functional role in improving neural retrieval.
\end{itemize}

\section{Related Work}

Our work connects neural information retrieval, query rewriting, and explainable IR. The key gap we address is that query rewriting and explainability are not connected in a feedback loop: rewrites are  produced without signals from the target retriever, and explanations are reported after retrieval rather than used to guide edits. \\[0.15cm]
\textbf{Neural IR} includes (i) expensive interaction models (e.g., cross-encoders) that jointly encode query--document pairs and are mainly used for re-ranking, and (ii) efficient retrieval models that encode queries and documents separately into sparse or dense representations for large-scale search (e.g., dual encoders and hybrid variants) \cite{nogueira2019bert,nogueira2020passage,karpukhin2020dense,Thakur2024}. Late-interaction architectures, such as ColBERT~\cite{khattab2020colbert}, retain token-level matching while remaining scalable, and hybrid sparse--neural methods such as SPLADE~\cite{spladev2} combine lexical matching with learned expansion \cite{khattab2020colbert,tctcolbert,spladev2}. Our approach is compatible with these efficient retrievers and focuses on improving retrieval without changing the retriever itself.\\[0.15cm]
\textbf{Query reformulation }has long been used to improve retrieval, from relevance feedback and pseudo-relevance feedback to modern neural and generative approaches \cite{robertson1990term,harman1992relevance,nogueira2017task,ren2018generating}. Recently, LLMs have enabled fluent query rewrites in zero-/few-shot settings and have shown gains for short or under-specified queries \cite{raffel2020exploring,jia-etal-2024-mill,jagerman2023queryexpansionpromptinglarge,10.1145/3748304}. However, most neural and LLM-based rewriting methods optimize for linguistic plausibility or semantic similarity without explicitly using signals from the downstream retriever, which can lead to inconsistent gains when the retriever is sensitive to query phrasing \cite{khattab2020colbert}. In contrast, we guide rewriting using token-level signals derived from the retriever’s scoring behaviour.\\[0.15cm]
\textbf{Explainable IR} aims to identify which query or document components drive neural retrieval decisions. Prior work explored attention-based analyses, gradient-based attributions (including Integrated Gradients), and perturbation-based approaches that measure the effect of masking query terms \cite{11272475,ribeiro2016should,sundararajan2017axiomatic,boytsov2021exploring,zhan2020analysis}. These methods are valuable for analysis and debugging, but they are usually treated as post-hoc explanations rather than integrated into the retrieval pipeline \cite{10849387,han2021explaining}. 
Our work operationalizes these explanation signals by using token-level attributions as guidance for query rewriting.

\section{Attribution-Guided Query Rewriting}
\label{sec:method}

When a neural retriever returns weak results, the issue often lies in its sensitivity to the query’s surface form: some tokens steer scoring toward relevant documents, while others are down-weighted or steer retrieval toward irrelevant regions of the corpus. 
Our objective is to improve retrieval through targeted query reformulation: we aim to disambiguate or clarify the parts of the query that confuse the retriever while preserving its original intent. Importantly, the retriever itself is kept fixed, as retraining or modifying deployed systems is frequently impractical, making query-time reformulation the most viable intervention. To this end, we use explanation signals from the retriever as a lightweight control signal for rewriting. We first estimate the contribution of each query token to the retriever's scores and then guide an LLM to rewrite the query using these token-level attributions as soft guidance. The method is retriever-aware and model-agnostic.  


\subsection{Problem Formulation}
Let $q = \{t_1, t_2, \dots, t_n\}$ be an input query composed of a sequence of tokens, and let $\mathcal{D}$ denote a collection of documents. A neural retriever $R$ assigns a relevance score $s(q, d)$ to each query-document pair and produces a ranked list of documents.

Given an initial query $q$, the objective is to generate a rewritten query $q'$ such that the retrieval effectiveness under the same retriever $R$ is improved, without modifying or retraining $R$ . This design targets realistic deployment settings where the retriever may be fixed (e.g., proprietary, costly to retrain, or lacking supervision), and where query-time interventions are the most practical lever for improvement. The rewriting process is constrained to preserve the user's intent, while addressing query components that negatively affect retrieval performance because they are  weak, ambiguous, or misleading for this retriever. Unlike generic query rewriting strategies~\cite{nogueira2017task,jia-etal-2024-mill,10.1145/3748304}, we explicitly rely on token-level attribution signals derived from the retriever to guide the reformulation process.

\subsection{Attribution-Guided Prompting for Query Rewriting}
A key requirement is to obtain a token-level signal that reflects the retriever's current scoring behaviour.
For each query $q$, we first retrieve the top-$k$ documents $D_q^{top}$ using the base retriever $R$. 
We use these top-ranked documents because they represent what the retriever currently considers most relevant for $q$; attributing against them provides a query-token diagnostic aligned with the retriever's actual behaviour on this input.
We compute token attributions using Integrated Gradients (IG)~\cite{lundstrom2022rigorous} with respect to the relevance score $s(q, d)$ for each document $d \in D_q^{top}$.  
IG provides a fine-grained estimate of how much each query token contributes to the score under the retriever's internal representations.
To obtain a single robust score per token, we aggregate attributions across the top-$k$ documents:
\vspace{-0.2cm}
\begin{equation}
    \alpha_i = \frac{1}{k} \sum_{d \in D_q^{top}} \text{IG}(t_i, s(q,d))
\end{equation}

\vspace{-0.2cm}
\noindent where $\text{IG}(t_i, s(q,d))$ measures the contribution of token $t_i$ to the relevance score of document $d$. 
Averaging across documents stabilizes the signal: a token may be helpful for matching some retrieved documents but harmful for others, and we want a query-level estimate of its overall utility under $R$.
Tokens with low or negative $\alpha_i$ are considered as candidates for clarification or disambiguation as they contribute weakly to the current retrieval scores or may steer retrieval toward irrelevant content.

\subsection{LLM-Based Query Reformulation}
Attribution identifies which parts of the query matter to the retriever, but it does not specify how to repair them (e.g., expand an abbreviation, add specificity, or resolve an ambiguity). We therefore use an LLM as a flexible rewriting module, while constraining it with retriever-derived token signals.

All query tokens $t_1, \dots, t_n$ with their attribution scores $\alpha_1, \dots, \alpha_n$ are passed to the LLM, 
together with the original query $q$ and an instruction to reformulate it for improving retrieval while preserving user's intent. 
The full prompt is shown listing 1.
We do not remove low-attribution tokens before rewriting. Low attribution does not mean that a token is irrelevant to the user intent; it may encode context or a constraint that the retriever currently fails to exploit. 
Instead, we treat attribution as soft guidance: the LLM is encouraged to preserve or emphasize high-attribution tokens and to clarify low- or negative-attribution tokens (e.g., by making them more specific or less ambiguous), while avoiding the introduction of new concepts.

\lstset{style=mystyle}
\label{list1: prompt}
\begin{lstlisting}[language=Python, caption=Prompt for attribution-guided query rewriting.]
prompt = """You are given:
1) An original user query.
2) A list of query tokens with their attribution scores, where higher scores indicate a stronger positive contribution to retrieval effectiveness, and lower or negative scores indicate weak or misleading contributions.

Your task is to rewrite the query to improve retrieval effectiveness.

Guidelines:
- Preserve the original user intent.
- Do not remove important concepts.
- Tokens with high attribution scores should be preserved or emphasized.
- Tokens with low or negative attribution scores may be clarified, specified, or disambiguated.
- Avoid adding new concepts that are not implied by the original query.
- Produce a single rewritten query, concise and well-formed.

Original query: "{original_query}"
Token attributions: "{zip(tokens, attributions)}" """
\end{lstlisting}


The LLM output rewritten query $q'$ is then passed to the same neural retriever $R$ to obtain a new ranked list of documents. We operate in a one-shot closed-loop setting (one attribution step and one rewrite per query), 
which keeps the method efficient and avoids propagating errors across multiple rewriting rounds.
The entire method operates without modifying the retriever’s parameters and does not require additional training data or relevance judgments.

\section{Experimental Setup}
\label{sec:experiments}

\sloppy
\subsection{Datasets,  Retrievers, and LLM}
We use the standard neural information retrieval benchmark BEIR~\cite{thakur2021beir}, a heterogeneous set of 18 IR datasets covering multiple domains, including scientific, news, and biomedical collections. Here, We focus on datasets covering both open-domain and domain-specific retrieval settings: \textit{SciFact} (Scientific), \textit{FiQA-2018}(Finance), and \textit{NFCorpus}(Bio-Medical). SciFact consists of 5K documents and 300 queries with 12.37 average word length per query (words/query), whereas FiQA-2018 includes 57K documents and 648 queries with 10.77 words/query, and NFCorpus comprises 3.6K documents and 323 queries averaging 3.30 words/query.

We consider two representative neural retrieval approaches: hybrid sparse-dense  retrieval and contextualized dense retrieval. For the first, we use SPLADE~\cite{spladev2}, which combines sparse lexical matching with learned semantic representations through query and document expansion. For the second, we use TCT-ColBERT~\cite{tctcolbert}, which captures fine-grained token-level interactions while enabling efficient retrieval through compressed dense representations.

Late-interaction dense retrievers such as TCT-ColBERT, trained on web-scale QA collections like MS MARCO, exhibit limited zero-shot generalization on heterogeneous datasets like NFCorpus and FiQA-2018~\cite{thakur2021beir}. Therefore, we fine-tune TCT-ColBERT models on the studied datasets prior to the experimental study.

For query rewriting, we use  Mistral (See Implementation details). 

\vspace{-0.15cm}

\subsection{Baselines and Evaluation Metrics}
We compare our \textbf{Attribution-guided LLM rewriting (GLLM}) method  against the following baselines -some of which play the role of ablation studies:

\begin{itemize}
    \item \textbf{Original Query (Org in Table~\ref{tab:main_results})} where the original query is directly processed by the neural retriever.
    \item \textbf{LLM-only Rewriting (LLM)} where queries are rewritten using the LLM model and prompt from  Listing\,1, but without incorporating any token-attribution information.
     \item \textbf{Top-Attribution Tokens Only (Tkn)} uses a reduced query containing only tokens with attribution scores above the query-wise mean, assessing whether highly contributive tokens alone suffice for effective retrieval.
\end{itemize}

We adopt standard retrieval metrics to evaluate ranking effectiveness at multiple cutoff values $k \in \{1,3,5,10,100\}$: nDCG@$k$, MAP@$k$, and Precision@$k$.
All metrics are computed per query and then averaged over the evaluation set.

\vspace{-0.2cm}

\subsection{Implementation Details}

Token-level attributions are computed using \textit{Integrated Gradients} with respect to the retriever relevance score. For each query, attributions are calculated against the top-$k$ retrieved documents, with $k=5$ in all experiments, and then averaged to obtain a single attribution score per query token -we keep the impact of $k$ for future work. Attribution scores are normalized on a per-query basis.

All query tokens, together with their corresponding attribution scores, are provided to the \texttt{Mistral-7B-Instruct-v0.3}-based language model accessed via the Hugging Face Transformers library\footref{mistral}. We use the prompt in Listing\,1 that includes the original query, the full list of tokens with their attribution values, and explicit instructions to reformulate the query. 
No tokens are removed prior to rewriting; attribution values serve solely as guidance for the reformulation process. Query rewriting is performed in a one-shot manner using greedy decoding. The length of the generated query is constrained to a maximum of 120 tokens. 
The source code for all experiments is publicly available on GitHub\footnote{\url{https://github.com/anonym-submission-code/Attribution-Guided-Query-Rewriting-for-Neural-Information-Retrieval}}.

\begin{table*}[h!]
\caption{ Retrieval effectiveness for original queries\,(Org), top-attribution tokens only\,(Tkn), LLM-only rewriting\,(LLM), and attribution-guided LLM rewriting (GLLM). Best results are in bold -blue. }
\centering
\small 
\setlength{\tabcolsep}{2pt} 
\renewcommand{\arraystretch}{0.9}
\begin{tabular}{l|l|cccc|cccc|cccc||cccc|cccc|cccc}
\multirow{3}{*}{\rotatebox{90}{Collection}}
 &\multirow{4}{*}{\quad\textbf{$k$}} & \multicolumn{12}{c||}{\textbf{SPLADE}}& \multicolumn{12}{c}{\textbf{TCT-ColBERT}} \\
\cmidrule(r){3-26}
& & \multicolumn{4}{c|}{\textbf{NDCG}} 
 & \multicolumn{4}{c|}{\textbf{MAP}} 
 & \multicolumn{4}{c||}{\textbf{P}} 
 & \multicolumn{4}{c|}{\textbf{NDCG}} 
 & \multicolumn{4}{c|}{\textbf{MAP}} 
 & \multicolumn{4}{c}{\textbf{P}} \\
 \cmidrule(r){3-26}
 & & Org & Tkn & LLM & GLLM & Org & Tkn & LLM & GLLM & Org & Tkn & LLM & GLLM & Org & Tkn & LLM & GLLM & Org & Tkn & LLM & GLLM & Org & Tkn & LLM & GLLM \\
\midrule
\multirow{5}{*}{\rotatebox{90}{Nfcorpus}}
 & @1  & .433 & .260& .394& \color{blue}\textbf{.461}& .055& .026& .046& \color{blue}\textbf{.059}& .442& .272& .402& \color{blue}\textbf{.476}& .444& .249& .373& \color{blue}\textbf{.476}& .054& .023& .045& \color{blue}\textbf{.056}& .464& .263& .393& \color{blue}\textbf{.495}\\
 & @3  & .392 & .250& .357& \color{blue}\textbf{.398}& .095& .052& .081& \color{blue}\textbf{.099}& .368& .244& .332& \color{blue}\textbf{.375}& .416& .231& .354& \color{blue}\textbf{.428}& .095& .046& .075& \color{blue}\textbf{.098}& .401& .229& .345& \color{blue}\textbf{.407}\\
 & @5  & .366 & .244& .327& \color{blue}\textbf{.390}& .108& .063& .092& \color{blue}\textbf{.113}& .312& .220& .277& \color{blue}\textbf{.339}& .407& .227& .348& \color{blue}\textbf{.414}& .117& .055& .093& \color{blue}\textbf{.120}& .367& .210& .321& \color{blue}\textbf{.369}\\
 & @10 & .335 & .228& .298& \color{blue}\textbf{.364}& \color{blue}\textbf{.127}& .073& .107& .123& .244& .177& .218& \color{blue}\textbf{.248}& .390& .228& .339& \color{blue}\textbf{.394}& .149& .071& .120& \color{blue}\textbf{.152}& \color{blue}\textbf{.313}& .190& .279& \color{blue}\textbf{.313}\\
& @100 & \color{blue}\textbf{.296} & .221& .268& .295& \color{blue}\textbf{.154}& .097& .131& .151& .074& .070& .066& \color{blue}\textbf{.079}& .399& .254& .353& \color{blue}\textbf{.401}& .227& .120& .189& \color{blue}\textbf{.229}& \color{blue}\textbf{.116}& .086& .106& .114\\

\midrule
\multirow{5}{*}{\rotatebox{90}{FiQA-2018}}
 & @1  & .333& .214& .266& \color{blue}\textbf{.337} & .173& .113& .144& \color{blue}\textbf{.178} & .333& .214& .266& \color{blue}\textbf{.338} & .345& .158 & .290 & \color{blue}\textbf{.382} & .175 & .080 & .147 & \color{blue}\textbf{.196} & .345 & .158 & .290 & \color{blue}\textbf{.382} \\
 & @3  & .298& .209& .252& \color{blue}\textbf{.308} & .233& .158& .197& \color{blue}\textbf{.236} & .190& .136& .160& \color{blue}\textbf{.195} & .319 & .157 & .279 & \color{blue}\textbf{.328} & .243 & .117 & .213 & \color{blue}\textbf{.254} & .211 & .105 & .184 & \color{blue}\textbf{.218} \\
 & @5  & .314& .221& .261& \color{blue}\textbf{.321} & .253& .173& .211& \color{blue}\textbf{.255} & .142& .103& .116& \color{blue}\textbf{.147} & .329 & .169 & .294 & \color{blue}\textbf{.336} & .262 & .130 & .231 & \color{blue}\textbf{.272} & \color{blue}\textbf{.154} & .084 & .140 & .150 \\
 & @10 & .341& .244& .288& \color{blue}\textbf{.337}& .269& .186& .227& \color{blue}\textbf{.278} & .091& .069& .077 &  \color{blue}\textbf{.099} & .356 & .192 & .324 & \color{blue}\textbf{.366} & .282 & .142 & .250 & \color{blue}\textbf{.292} & \color{blue}\textbf{.099} & .058 & .091 & \color{blue}\textbf{.099} \\
 & @100 & \color{blue}\textbf{.398}& .306& .351& .389 & \color{blue}\color{blue}\textbf{.284}& .201& .241& .280 & \color{blue}\textbf{.015}& .013& .014& .013 & .424 & .265 & .394 & \color{blue}\textbf{.433} & .299 & .157 & .267 & \color{blue}\textbf{.309} & .016 & .013 & .016 & \color{blue}\textbf{0.17} \\

\midrule
 \multirow{5}{*}{\rotatebox{90}{SciFact}}
 & @1  & .560 & .190 & .546 & \color{blue}\textbf{.603} & .530 & .172 & .524 & \color{blue}\textbf{.572} & .560 & .190 & .546 & \color{blue}\textbf{.603} & .479 & .023 & .456 & \color{blue}\textbf{.526} & .479 & .020 & .442 & \color{blue}\textbf{.502} & .503 & .023 & .456 & \color{blue}\textbf{.526 }\\
 & @3  & .632 & .255 & .612 & \color{blue}\textbf{.654} & .604 & .231 & .588 & \color{blue}\textbf{.630} & .247 & .117 & .235 & \color{blue}\textbf{.250} & .558 & .044 & .559 & \color{blue}\textbf{.602 }& .558 & .038 & .527 & \color{blue}\textbf{.573} & .234 & .021 & .224 & \color{blue}\textbf{.235} \\
 & @5  & .653 & .286 & .635 & \color{blue}\textbf{.676} & .619 & .250 & .604 & \color{blue}\textbf{.646} & .162 & .088 & .156 & \color{blue}\textbf{.165} & .567 & .051 & .571 & \color{blue}\textbf{.615} & .567 & .042 & .535 & \color{blue}\textbf{.582} & .150 & .016 & .142 & \color{blue}\textbf{.151} \\
 & @10 & .677 & .309 & .660 &\color{blue}\textbf{ .696} & .631 & .260 & .616 &\color{blue}\textbf{ .656} & .090 & .052 & .087 &\color{blue}\textbf{ .090} & .577 & .066 & .598 & \color{blue}\textbf{.636} & .577 & .048 & .548 & \color{blue}\textbf{.592} &\color{blue}\textbf{ .083} & .0133 & .082 & \color{blue}\textbf{.083 }\\
 & @100 & .705 & .369 & .693 &\color{blue}\textbf{ .724} & .637 & .271 & .624 &\color{blue}\textbf{ .662} & \color{blue}\textbf{.010} & .007 & \color{blue}\textbf{.010} & \color{blue}\textbf{.010} & .585 & .122 & .631 & \color{blue}\textbf{.668} & .585 & .058 & .555 & \color{blue}\textbf{.600 }& \color{blue}\textbf{.010} & .004 & \color{blue}\textbf{.010} & \color{blue}\textbf{.010} \\

\midrule
\end{tabular}
\label{tab:main_results}
\end{table*}

\section{Results and Analysis}
\label{sec:results}

Table~\ref{tab:main_results} summarizes the effectiveness for SPLADE, TCT-ColBERT using nDCG, MAP, and Precision at $k \in \{1,3,5,10,100\}$. Results consistently show that the proposed attribution-guided query rewriting approach outperforms the original query and all baseline rewriting strategies (similar to ablation studies) across models and cutoffs.  

\subsection{Comparison with Baselines}

The Top-Attribution Tokens Only baseline (Tkn column) 
performs worst overall: while it emphasizes highly contributive terms, it consistently underperforms both the original queries  (Org) and the LLM-based rewriting approaches. 
Across all retrievers, LLM-based rewriting without attribution guidance (LLM) yields limited and inconsistent gains, sometimes even degrading retrieval effectiveness compared to using the original query (Org column). 
In contrast, attribution-guided LLM rewriting (GLLM) consistently improves retrieval effectiveness, with NDCG@10 gains of up to 9\% over original queries and up to 22\% over LLM-only rewriting for SPLADE, and up to 10\% and 16\%, respectively, for TCT-ColBERT. 
All together, these results indicate that low-attribution tokens, while individually weak, often provide important contextual cues that are critical for effective retrieval.

\vspace{-0.1cm}
\subsection{Qualitative Analysis}

To better understand the behaviour of the model, we analyse the rewritten queries produced by our attribution-guided method\,(see an example Table~\ref{tab:qualitative}). Tokens with high positive attribution scores are generally preserved or emphasized, whereas ambiguous or weakly contributive tokens are clarified or disambiguated. For instance, abbreviated or underspecified terms are expanded, and vague expressions are replaced with more precise alternatives. In contrast, LLM-based rewriting without attribution signals occasionally introduces unnecessary paraphrases or semantic drift. This qualitative difference explains the consistent quantitative gains observed with attribution-guided reformulation.

\begin{table}[h]
\centering
\caption{Example of query rewriting using baseline and attribution-guided methods.}
\label{tab:qualitative}
\begin{tabular}{p{0.25\linewidth} p{0.67\linewidth}}
\toprule
\textbf{Original query} &
What is actually in chicken nuggets? \\
\midrule
\textbf{Tokens (attrib.)} &
\texttt{what}~(0.008), 
\texttt{is}~(0.012), 
\texttt{actually}~(0.013), 
\texttt{in}~(0.018), 
\texttt{chicken}~(0.217), 
\texttt{nuggets}~(0.648) \\
\midrule
\textbf{Top Tokens} &
chicken nuggets \\
\midrule
\textbf{LLM-based} &
What is in chicken nuggets? \\
\midrule
\textbf{LLM-guided} &
What are the ingredients in chicken nuggets? \\
\bottomrule
\end{tabular}
\vspace{-0.2cm}
\end{table}

Overall, the results show that incorporating token-level attribution into query rewriting leads to consistent and significant improvements in neural retrieval effectiveness. The analysis shows that attribution signals are more effective when used as soft guidance for reformulation rather than as a hard filtering mechanism. 

Importantly, the extraction of token-level attributions incurs negligible computational overhead, as it is performed directly within the standard retrieval process. 
Further results, detailed efficiency analyses, and extended interpretations are provided in the supplementary material available in the associated GitHub repository.

\vspace{-0.5em}
\section{Conclusion}
\label{conclusion}

We present an attribution-guided query rewriting method that leverages token-level contributions from neural retrievers to inform LLM-based reformulation. By providing query tokens along with their attribution scores, our method selectively refines ambiguous or low-utility terms while preserving user intent. Experiments on SPLADE and TCT-ColBERT across BEIR collections demonstrate consistent improvements in nDCG, MAP, and Precision.
Our results highlight that attribution-guided query rewriting enables more effective query reformulation than naive LLM rewriting or token pruning, improving early ranking and semantic alignment without retriever retraining. The method is model-agnostic, efficient, and compatible with multiple neural retrieval paradigms. 
In future work, we plan to explore additional neural retrievers and benchmark collections, investigate iterative and multi-stage rewriting strategies, and assess cross-lingual and domain-specific retrieval scenarios. We also aim to study alternative attribution methods and their interaction with large language models.

\bibliographystyle{ACM-Reference-Format}
\bibliography{bibliography}


\end{document}